\documentclass[preprint,
nofootinbib,
preprintnumbers,
amsmath,amssymb]{revtex4-2}

\newtheorem{mydef}{Definition}

\newtheorem{myhyp}{Hypothesis}

\newcounter{hipotesis}

\newcommand{\hipotesis}[2]
{\refstepcounter{hipotesis}%
\par\noindent 
{\bf Hipótesis} \thehipotesis\ 
[#1]:
 \, #2\medskip\par}

\usepackage{graphicx,epsfig}
\usepackage{simplewick}
\usepackage{slashed}
\usepackage{color}
\usepackage{setspace}
\usepackage{hyperref}
\usepackage{empheq}

\usepackage[utf8]{inputenc}

\def\simg{{\ \lower-1.2pt\vbox{\hbox{\rlap{$>$}\lower6pt\vbox{\hbox{$\sim$}}}}\ }}
\def\siml{{\ \lower-1.2pt\vbox{\hbox{\rlap{$<$}\lower6pt\vbox{\hbox{$\sim$}}}}\ }}

\newcommand{\be}{\begin{equation}}
\newcommand{\ee}{\end{equation}}
\newcommand{\bea}{\begin{eqnarray}}
\newcommand{\eea}{\end{eqnarray}}

\def\siml{{\ \lower-1.2pt\vbox{\hbox{\rlap{$<$}\lower6pt\vbox{\hbox{$\sim$}}}}\ }}

\def\simg{{\ \lower-1.2pt\vbox{\hbox{\rlap{$>$}\lower6pt\vbox{\hbox{$\sim$}}}}\ }}
\def\siml{{\ \lower-1.2pt\vbox{\hbox{\rlap{$<$}\lower6pt\vbox{\hbox{$\sim$}}}}\ }} 

\begin{document}

\vskip 1truecm
\title{Relativity: A matter of causality}
\author{Antonio Pineda}
\address{
{\small ${}^1$ \it Grup de Física Teòrica, Dept. Física,}\\
{\small \it Universitat Autònoma de Barcelona, E-08193 Bellaterra, Barcelona, Spain}
\\
{\small ${}^2$ \it Institut de Física d'Altes Energies (IFAE),}
{\small\it The Barcelona Institute of Science and Technology,}\\
{\small\it Campus UAB, 08193 Bellaterra (Barcelona), Spain}}
\begin{abstract}
We take causality and uniqueness of events observation as our driving forces. They are built in in the way we define distinct observers, which then require a finite time to communicate between each other. This unavoidably leads to the existence of maximal transfer-information velocity between arbitrary (not necessarily inertial) reference frames. Inertial reference frames are defined by fixing the geometrical properties of (spatial) distance without any reference to relativity, electromagnetism, or laws of physics in general. For these inertial reference frames, the causality condition fixes the causal group to be the orthochronous inhomogeneous Lorentz group times dilatations. The mathematics we will use are quite basic.
\end{abstract}


\maketitle

\vfill
\newpage

\tableofcontents

\newpage        

\pagenumbering{arabic}

\begin{flushright}
{\bf "Observo, luego existo"}\\
\end{flushright}

\section{Introduction}

Since the seminal derivation of special relativity by Einstein \cite{Einstein:1905ve}, based on the constancy of the speed of light and the relativity of inertial reference frames, there have been many alternative ways to try to get the same (or similar) results based on alternative sets of assumptions (see \cite{Ignatowsky,LeeKalotas,Leblond,Pelissetto:2015bva,CornellaLatorre,LlosaMolina} for a particular selection of them). The most common motivation to seek for alternative derivations  is the special role played by light in these derivations, whereas special relativity is thought to be a more general setup, not linked to the theory of electromagnetism. Indeed, it was soon realized \cite{Ignatowsky} (see \cite{LeeKalotas,Leblond,Pelissetto:2015bva} for more recent related work) that Lorentz-like transformation rules could be obtained assuming homogeneity of space-time and isotropy of space, plus the condition of relativity between inertial reference frames. Another derivations of Poincare symmetry demand the invariance of the space-time interval under transformations between inertial reference frames. Several other possible combinations of hypotheses also exist (see, for instance, \cite{CornellaLatorre,LlosaMolina}).

We do not aim here to give a full account of all different ways to address special relativity. Instead, here we would like to take a different approach compared with most derivations of special relativity aimed to undergraduate students. There are several reasons for that:
\begin{enumerate}
\item
Certainly, we would like to avoid using light/electromagnetism as one of the basic principles for the construction of special relativity, as we share the view that it should be possible to get special relativity using more general arguments.
\item
Reference frames and inertial reference frames are often defined in a somewhat {\it vague way} using statements such that: "the laws of physics" ... "take a simpler form"/"are invariant" .. .."when no forces act on the particles". This is unsatisfactory: What are the laws of physics? What does it mean "simpler form"? What does it mean "invariant"? What are forces? ...). All these statements use concepts that have not usually been defined at the level of the derivation of special relativity. This is completely avoided in our derivation.
\item
Whereas the concept of relativity between reference frames is very intuitive, it is indeed possible to derive special relativity without using it (at least in the way it is usually implemented). 
\end{enumerate}

Compared with the points above our stands is the following. 
We take causality and uniqueness of events observation as our driving forces. They are built in in the way we define distinct observers, which then require a finite time to communicate between each other. This unavoidably leads to the existence of a maximal transfer-information velocity between arbitrary (not necessarily inertial) reference frames. These are carefully defined in a constructive way. Inertial reference frames are a particular subset of them, being defined by fixing the geometrical properties of (spatial) distance. This definition does not make any reference to relativity, electromagnetism, forces, or laws of physics in general. For these inertial reference frames, the causality condition fixes the causal group to be the orthochronous unhomogeneous Lorentz (Poincaré) group times dilatations \cite{Alexandrov,Zeeman} if the number of spatial dimensions, $d$, is greater than one.

The implementation of these assumptions will be made in such way that the theory is internally logically consistent. In other words, it does not lead to paradoxes (in mathematical terms, one would say that one is using {\it reductio ad absurdum} wherever necessary).

Whereas the mathematics we will use are quite basic (when they get complicated, we will refer to results from the literature), the presentation has a mathematical structure that can make it more appealing to first year students of a physics course in the Mathematics degree, or to students of the physics degree more oriented to theoretical physics.  

\section{Definitions}

\subsubsection{Time}
\noindent
{\bf Time}: \underline{order relation}. Humans have the ability to discern between \underline{before}, \underline{after}, and \underline{same}  (or organize the description of nature in this way).
\\
{\bf Clock}. Mechanism/object that is \underline{always} changing (there is dynamics here). We can choose the change to be in a closed cycle. The clock will have a mechanism to count cycles. This is what we call \underline{time}. Time is a strictly {\bf increasing} function.\\
{\it Observation}. The clock carries with it the concept of \underline{change} in a continuous and eternal way. 
We want this change to be uniform (the magnitude of the cycle does not depend on when the clock started working).   
This is an idealization, it implicitly carries the concept of conservation of energy. A through discussion can be found in \cite{Poincare} (see also \cite{CesarGomez} for a nice talk dwelling on this issue). The idea behind is that the clock is a closed system not interacting with the environment. Strictly speaking this is not true when we look to the clock, since then there is some interaction. This is particularly relevant at the quantum level. Still, if the interaction of the clock with the environment is very small, one can still consider the clock to be ideal, and the interaction to be considered as a small perturbation one accounts for as an external force. Other than this observation, we will not dwell further on this issue in the derivation of relativity presented in this work, and consider that we have a set of idealized clocks. 
\\
{\it Observation}. Example of a clock. An ideal pendulum, and counting the number of cycles. At the atomic level, one can think of the oscillation time between states, or, when working with electromagnetic classical fields, one may measure the frequency change of its associated amplitude over time. Typically, this discretizes time. In a classical pendulum, one might think of measuring a fraction of the cycle. If, for each cycle, we can measure the fraction, we could make time continuous. This implies doing measurements at different positions (if we only measure the complete cycle we are measuring when the pendulum is in the same place, and there is no need to consider different positions). This problem, in principle, does not show up in the case of a classical electromagnetic field. Still, these kind of considerations can be problematic at the quantum level. We will ignore all these issues and consider we are working with idealized systems for which the time intervals can be made arbitrarily small.  \\

\medskip

\subsubsection{Observer/emitter (OE)}
\noindent
{\bf OE}: A person with a \underline{clock} and eyes + any other measuring device: Stern-Gerlach, .... It can also be an emitter. It can emit rays, reflect them, generate electromagnetic or gravitational waves, etc.... In general, it can be considered as a set of measuring devices with a clock located at a point in space.\footnote{The definition of ``a point in space'' will be given when we talk of a coordinate representation of an OE.} The clock defines a time $t_i$ associated with the OE$_i$. This is the proper time of OE$_i$.

\medskip
\noindent
{\it Observation}. Time/clock is a "local" concept, associated with a point in space.

\noindent
{\bf Event} (associated with an OE): An observation or an action (emission ....) of the OE at a time $t$ of the OE clock.\\
{\it Observation}. This is an idealization, since the event may need a finite amount of time to take place. Part of this  idealization can be taken away if one defines $t$ as the initial time of the emission, and the remaining effect of the OE on the emitted object is included in the effects of the external medium into the emitted object.\\
{\it Observation}. This definition of event also assumes that once the event has finished, the outcome (we can think of an emitted particle) does not interacts back (directly) with the OE (the interaction is local). \\
{\it Observation}. Note that the concept of {\bf event} combines {\bf OE} (which eventually will be space) and time.
\\
{\it Observation}. An event is a binary process. It either happens or it doesn't. We will assign numerical values to these events. The observation/event can always be \underline{numerized}.\\
{\it A follow up of this observation} is that the result of the observation/event is then \underline{unique} (this is a key property that defines what a \underline{function} is, as emphasized by Dirichlet in the XIX century). By the very nature of observation, we only get one result.\footnote{At the quantum level, this may lead to thinking of paradoxes. These will only happen if we assign reality to what we do \underline{not} observe. The only thing that has to be \underline{unique} is the observation, nothing else. Given an observer, an event occurs or does not (here one could worry about what the threshold to detect the event is, but, given an OE, the situation is binary).}
\\
{\it Examples}: Assign a numeric value 1 if a certain event occurs and 0 if it does not. This defines a function of $t$. Obviously, much more sophisticated numerical characterizations of the events can be devised, but they will still be functions of $t$. We name these functions of $t$  {\bf Observables}:
\be
O_i(t_i).
\ee
This quantity represents the numerical value obtained for the observable $O$ by the $OE_i$ at the time $t_i$.

The clock and the observables define completely the OE. Being more precise:
\\
{\bf Definition}: One {\bf OE$_i$} is characterized (defined) by a {\bf local time $t_i$} and by a set $\{\alpha\}$  of {\bf observables $O_i^{\{\alpha\}}(t_i)$}.
\\
{\bf Simultaneity of an OE}: When two events happen at the same \underline{now} (time) of the observer.

\medskip

We now consider a second OE: OE$_j$. We assume the observables (experimental apparatus) to be equal. This means that we have callibrated the observables at an equal place and then moved to a different place. All observers can measure and emit the same things: $O_j^{\{\alpha\}}(t_j)$. 

\subsubsection{Messengers}
\noindent
\begin{mydef}\label{def:messengers}
{\bf Messengers}: Methods of information transfer between OEs.
\end{mydef}

\noindent
We now give examples, and show how they can potentially \underline{define} distances between OEs:\\
1) The OE$_i$ launches an arrow with a well-calibrated mechanism (it does not wear out) when OE$_i$'s clock marks $t_i^{(in)}$. When it reaches the observer OE$_j$, it cuts a string that launches a return arrow that reaches back the observer OE$_i$ when its clock marks $t_i^{(final)}$. Then, we define the {\bf distance} from OE$_j$ to OE$_i$ associated with this method $(a)$ as
\be
d_{ij}^{(a)}(t_i)\equiv \frac{t^{(final)}_i(t_i^{(in)},(a),j)-t_i^{(in)}}{2}
\,.
\ee 

2) Each OE$_i$ has a catapult system. Each catapult has a person with an infinitely sharp knife. OE$_i$ activates the catapult and launches the person with an infinitely sharp knife when its clock reads $t_i^{(in)}$. When it reaches the OE$_j$, it cuts a rope that activates another catapult that launches another person back who reaches OE$_i$ when its clock reads $t_i^{(final)}$. We define the {\bf distance} between the two OEs using this method (c) as
\be
d_{ij}^{(c)}(t_i) \equiv \frac{t^{(final)}_i(t_i^{(in)},(c),j)-t_i^{(in)}}{2}
\,.
\ee
3) Each OE$_i$ has a system of flashlights and mirrors. The OE$_i$ turns on the flashlight when his clock reads $t_i^{(in)}$. When it reaches OE$_j$ it is reflected in the mirror. It returns to OE$_i$ when its clock reads $t_i^{(final)}$. We define the {\bf distance} between the two OEs using this method (l) as
\be
d_{ij}^{(l)}(t_i) \equiv \frac{t^{(final)}_i(t_i^{(in)},(l),j)-t_i^{(in)}}{2}
\,.
\ee

Overall, we define the {\bf distance}\footnote{At this point, we do not claim that this definition alone furnishes a distance (even if we name it distance), further input is needed.} {\bf from OE$_i$ to OE$_j$ associated with method $(m)$} as
\be
\label{def:dm}
d_{ij}^{(m)}(t_i)\equiv \frac{t^{(final)}_i(t_i^{(in)},(m),j)-t_i^{(in)}}{2}
\,.
\ee

{\it Observation}. So far, we have only considered the time $t_i$ of a single observer.

{\it Observation}. The concept of distance between observers requires we consider the OEs to be local. We will assume that this idealization does not generate logical problems in the theory.

{\it Observation}. For each messenger $m$, $t^{(final)}_i(m)$ will be different. The dependence on $t_i^{(in)}$ of $t_i^{(final)}(m)$ could be complicated.

{\it Observation}. There could exist a set of observers interacting with a disjoint set of messengers. We would never notice their presence. Therefore, we will ignore them following Occam's razor principle.

{\it Observation}. In principle, messengers could also be OEs (think of the catapult example before). 

\subsection{Causality and space}

{\bf Definition: Causally connected OEs}. We say two OEs are causally connected if there exists a set of messengers such that it is possible to define a distance between them using Eq. (\ref{def:dm}).

{\it Observation}. Note that this definition does not yet refer to causality of events, which is going to be discussed later.

\noindent
\begin{mydef}\label{def:distinct}
{\bf Two causally connected OEs are said to be located at different positions in space} during a finite period of time of the clock of the OE$_i$ if there is no messenger able to transmit information between OE$_i$ and OE$_j$ instantaneously during this OE$_i$ time interval.\\
Mathematically: $\exists \; i,j$ such that $d^{(m)}_{ij} >0 $ $\forall m$ for a OE$_i$ finite time interval.
\end{mydef}

{\it Corollary}. This definition is symmetric to the interchange of $i$ and $j$ ($d^{(m)}_{ij} >0 $ $\forall m$ $\iff$ $d^{(m)}_{ji} >0 $ $\forall m$), otherwise one would get contradictory results. 

\begin{myhyp}\label{def:Hp1}
{\bf There are in nature OEs located at different positions in space}. Actually, we will assume that we can generate a continuous family of them.
 \end{myhyp}

{\it Remark}. A given OE can be characterized in many ways by setting the experimental apparatus differently (Stern-Gerlach rotated, different frequency for the clocks, ....). This can be interpreted as the same OE in a different state (if using a quantum mechanics notation).  Alternatively, in some circumstances, it will also be convenient to consider these two different states as two different OEs but located at the same point in space.
                        
{\bf Corollary of hypothesis \ref{def:Hp1}. Causality of events}. Every event/action performed by the OE$_i$ at a time $t^{(in)}$ will be received back after interacting with OE$_j$ (with which it is causally connected) at a later time $t^{(final)} > t^{(in)}$ if OE$_i$ and OE$_j$ are located at different positions in space. In other words: every propagation in space of an action (relationship with another observer) requires a \underline{non-zero} finite time. This is in the theory by construction.

{\it Observation}. This corollary is "local", in the sense that it refers to the time of a single observer.

{\it Observation}. Let us emphasize again that, following how we have defined different OEs, \underline{causality} is not a hypothesis, it is a consequence of the definition of different OEs located at different positions in space, if anything, the hypothesis is the existence of different observers. Note that sending information is a particular case of ``action''. There is no method to transmit action (and in particular information) from one observer A to another observer B instantaneously. This is so by definition. If it could be done, we would be talking about the same observer or two observers at the same point in space. 

{\it Corollary of the causality of events}. If an event happens for one OE it also happens for any other OE with whom it can communicate. "They observe the same reality". This is the definition of an event happening for an OE who is not at the physical site of the event. This is nothing but the {\bf transitive property}, which will eventually lead us towards having a structure of groups in the relation of events as seen by different observers. Example: If something happens to an observer C, it sends a signal to A and B.

{\bf Definition of universal distance}.\footnote{This will be our default definition for distance, as it makes it independent of the specific messenger used.} By default, 
we define the distance as the minimum time to transfer information between different OEs:
\be
\label{distance}
d_{ij}(t_i) \equiv \underset{\{m\}}{\min}\{d_{ij}^{(m)}(t_i)\}
\,.
\ee
It is not compulsory, but we will assume that there is a non-zero set of messengers that saturate the minimum.\footnote{This is what seems to happen in nature.} Such definition will then single them out.  
 We will name them \underline{limit-messengers}. There may be more than one messenger for which its associated distance is equal to the minimum distance (this would mean that there exist some observables that allow us to distinguish between these messengers).
We will consider that there exists at least one limit-messenger that interacts with \underline{all} OEs.\footnote{This role is played in our world by the gravity interaction: The energy-momentum tensor for every particle is always nonzero. For the case of the photon, there are currents that are identically zero or, in other words, there are particles that do not interact with photons.}

{\it Remark}. Note again that, even if we name it distance, we have not yet shown that Eq. (\ref{distance}) fulfills all the properties the definition of distance has. 

{\bf Corollary of hypothesis \ref{def:Hp1}}. Given two observers located at different positions in space, $d_{ij}$ exists and is nonzero.

{\it Observation}. There could exist a set of observers that do not interact directly with the limit-messenger. This is not the case for us due to gravity. Nevertheless, let us consider such possibility anyhow. In such case, the propagation of information would be slower, but it could be made as close as we wanted to the case that interacts with the limit-messengers, by making that observer to interact with another observer that does indeed interact with the limit-messenger almost immediately. 

\subsection{Discussion: Instant interaction between OEs located at different points in space}
\noindent

Let us elaborate on the motivation for Definition \ref{def:distinct} and Hypothesis \ref{def:Hp1}. Our line of reasoning uses {\it reductio ad absurdum} arguments. Let us imagine that there is a method of transferring information that is instantaneous (instant interaction at a finite distance) but still have distinct observers. This would contradict our principle of causality/uniqueness of observation. By definition, such set of OEs would be in the same place (zero distance). We have eliminated such option by construction (we have made them to be the same OE, i.e. we have introduced an equivalence relation between these, potentially different, OEs), but why so? The reason is that their introduction may introduce logical loopholes. To illustrate the problem, let us emphasize that instantaneous interactions allow, formally, to do an infinite number of actions with $\Delta t=0$. Therefore, the OE could interact with itself at the same moment of doing the action. This opens the possibility that an event and its opposite could happen at the same instant. This is something that we forbid by construction. 

We can visualize this problem with the following example: Turn on a flashlight at A, and turn it off when we receive the response. At the same instant $t$ the flashlight would be on and off. This enters in contradiction/it is incompatible with demanding the result of an observation to be unique.     
In practice, we also demand that the flashlight also need a finite amount of time to change state, otherwise the observer runs in the same problem.   

There are legitimate concerns/questions that this discussion may rise. One is, what would the human eye/device see in this situation? Always on or always off? This, again, rises the question of whether introducing the concept of \underline{threshold} activation: Minimum energy to activate a process.
One might also think that the propagation is instantaneous but that the interaction with the receiver requires a finite time. In practice it would seem impossible to discriminate between both situations. Furthermore, experience seems to indicate that the speed is universal in all cases, whereas if it were dependent on the receiver one would expect receiver dependence. This, and other possible complications, we simply ignore and assume that we can work with the above mentioned hypotheses/definitions. Indeed, what it is certainly true is that such set of hypothesis/definitions creates a well defined mathematical structure with no internal loopholes.

Implicit in this discussion is that the interaction between different observers is always local. This means that, at a given time, both interacting observers/messengers are at the same point in space. It is only then than interaction takes place. 
         
\section{Universe}

\noindent
\begin{mydef}\label{def:Universe}
{\bf Universe}: The complete set of OEs: $A_i$, $i=1..N$ (we can consider a finite or infinite set of observers) that are  \underline{causally} \underline{connected}. Each one with their experimental apparatus to receive and send signals, and their clocks.  \end{mydef}

One may wonder whether one could consider the Universe as a reference frame. Addressing the problem this way turns out to be too complicated, and it is not the way we study physics. Instead, we rather consider the position of particles in space and how these positions change over time. Therefore, we have to give an operational meaning to (space) positions and time. This is better achieved considering the analogous to (three dimensional) rigid rods to label space and their associated clocks to label time. These are the rigid reference frames (RRFs) we will define in the next section. These RRFs will allow us to define a distance that fulfills the properties a distance definition has (therefore, we will have a metric space). In principle, each OE can generate an associated RRF (actually it is this generated RRF of each OE that allows us to quantify how events that happen in different positions of the OE are seen by the OE and how the outcome relates with the outcome of the event as seen by other OE). Then the causality principle constraints the allowed transformation relations between the different RRFs. This last item, we only discuss for RRFs with Euclidean metric. 

\section{Rigid reference frame (RRF)}

This is the (d-) three dimensional generalization of a rigid rod. 

\begin{mydef}\label{def:RRF}
{\bf RRF}: A set of OEs: $A_i$, $i=1..N$ (we can consider a finite or infinite set of OEs). Each one with their experimental apparatus to receive and send signals, and their clocks. We assume that these apparatus and clocks work in the same way if all OEs were at the same point in space, and that (note that $t_i^{(final)}$ also depends implicitly on $t_i^{(in)}$)
\end{mydef}
\be
\frac{d}{d t_i^{(in)}}d_{ij}(t_i^{(in)})=0 \quad \forall i,j
\;.
\ee 
This means that, irrespectively of when we emit the signal, we always obtain the same number.  This condition produces the following {\bf Corollary}:
\be
\label{dtm}
\frac{d}{d t_i^{(in)}}d_{ij}^{(m)}(t_i^{(in)})=0 \quad \forall i,j,m
\ee 
{\it Proof}: There always exists a OE$_n$ such that $d_{ij}^{(m)}=d_{in}+d_{nj}$. Since the right hand side is time independent we obtain Eq. (\ref{dtm}). 

\subsubsection{Clock synchronization within the same RRF}
\label{Sincr}
\noindent

So far, we have not synchronized the clocks of the different observers of the RRF. We do that in the following. 
The construction of the RRF has been made by taking a particular OE: OE$_i$ and its clock. Let's proceed to synchronize the clocks of the different OEs. We follow the method by Einstein \cite{Einstein:1905ve} with the only qualification of changing light by any limit-messenger.

For a given RRF, the distances between OEs are constant irrespectively of changing $i \leftrightarrow j$, since we can repeat the measurement process several times. We then first fix the frequency of the clocks of the observers $i$ and $j$ (the change pace of the hands of the clocks) by imposing 
\be
d_{ji}(t_j^{(in)}) =d_{ij}(t_i^{(in)}) \quad \forall i,j
\;.
\ee

In the second iteration, we send the value of $t_i^{(final)}(t_i^{(in)},(l),j)$. We synchronize the clock of $O_j$ ($t_j$) such that at the moment the limit-messenger (flash light) arrived the first time to $O_j$ we have 
\be
t_j=t_i^{(in)}+\frac{t_i^{(final)}(t_i^{(in)},(l),j)-t_i^{(in)}}{2}
\;
.
\ee
This procedure allows us to define a universal time, $t$, for all OEs of the RRF. 

{\bf Corollary}. Synchronization in these RRF have the transitivity property; If $A$ and $B$ are synchronized and $B$ and $C$ are synchronized, then $A$ and $C$ are synchronized. We prove this when we discuss geodesics and coordinate representations of observers and events. 

{\bf Remark}. We only synchronize the OEs clocks of the same RRF because it is in this situation that the distance is constant (independent of when we send the signal), and we can transmit the information of the result of A to B, so B can synchronize the result and the synchronization will remain for ever. This would not happen in general if there is relative movement between the OE's. The synchronization would depend on when it has been done. This discussion would be more relevant in general relativity and will not dwell on this here. 

\medskip

{\bf Complete rigid reference system}: RRF that occupies all space. \\
What does it mean that it occupies all space? That, at a given time of the RRF, for each OE of the universe, there exists a OE of the RRF such that they are at zero distance (but not necessarily the same OE if one takes a different time). It can be considered to be the limiting case of a generic RRF. One also considers the infinite timeline of one of the observers. Then the timeline of the other observers of the RRF is also infinity. It is useful as a mathematical idealization. Unless stated otherwise, we will assume the RRF expands over all space. 

\subsubsection{Geodesic Coordinates}

{\bf Coordinate representation of the different OEs of the RRF}\\
We are now in the position to define the coordinate axes. We arbitrarily take one OE and define it to be the OE$_{\bf 0}$, i.e. the OE located at ${\bf x}=0$ (the origin) by definition. The coordinates of the other observers can be defined by the speed-limit geodesics, i.e. by the minimal time needed for communication between two OEs.

{\bf One dimension}.
We first define one dimension. We first take one observer OE$_i$. We assign $i$ to be a positive integer number and name 
\be
x_i^1=d_{i0} 
\,.
\ee
We now consider all the observers, OE$_{i'}$,  that live in the geodesic, i.e. those for which there exist a combination that makes the triangular inequality to be saturated with the observers OE$_{\bf 0}$ and OE$_i$. There are three different possibilities:
\be
a) \; d_{i'i}+d_{i0}=d_{i'0}\;;
\qquad
b)\; d_{i i'}+d_{i' 0}=d_{i0} 
\;; \qquad c)\; d_{i0}+d_{0i'}=d_{ii'}
\,.
\ee
This produces an order relation in the set $\{\{i'\} \cup \{0\} \cup \{i\}\}$ (this set now includes 0 and $i$). In the first option $i' >i>0$, in the second $i>i'>0$ and in the third $i>0>i'$. 

We then define the associated coordinate to OE$_i$ and OE$_{\bf 0}$ as the set of OEs that live in the geodesic generated by OE$_i$ and OE$_{\bf 0}$, and label each element of this set by its distance to the origin times the sign of the label $i'$:
\be
 x_{i'}^1={\rm Sign}[i']d_{i'0} \;.
\ee
In the following we drop the label $i$, $i'$ and characterize the OE just by the number $x^1$: the label that characterize the whole set of OEs that live in the geodesic. 

{\it Remark}. Note that, by definition of RRF, $x^1$ is independent of $t$.

{\it Remark}. There is an order relation: $x^1$ is isomorphic to the one-dimensional real axis $\mathbb{R}$ (up to global considerations).

{\it Remark}. OE$_j$s that do not saturate the triangular inequality ($d_{j0}< d_{ji}+d_{i0})$) with $i$ and $0$ are outside the $x^1$ coordinate (they do not belong to this set).

Continuity and differentiability is associated to say that small variations of time produce small variations of distance. This follows from defining distance as a time difference, and taking time to be a continuous variable.

If for any set of 3 observers we have that there always exists a combination such that the triangular inequality is saturated we will have that the dimension is 1 (this can be understood as the definition of 1 dimension). If there exists any triad for which the inequality is strict the number of space dimensions is bigger than 1.

\medskip

{\bf More than one dimension}. We now consider that there are three different  points for which, if we send the limit-messenger from A, the geodesics AB and AC do not intersect (except in the point A). In physical terms, we can think that OE$_A$ sends two light pulses: one to OE$_B$ and the other to OE$_C$, and we find that the light does not go through both points $B$ and $C$.
This tells us that the number of space dimensions is bigger than one, as these points cannot be accommodated in the definition of one dimension. 
We take two of them ($A$ and $B$) to generate one dimension: $x^1$. Next, we look for $A'=$ \{the point of the $x^1$ coordinate that minimizes the distance with $C$\}:
\be
\underset{\{x^1\}}
{\rm min}\; d(C,x^1)=d(C,A')
\,.
\ee
Then $C$ and $A'$ define another geodesic, and, therefore, a new geodesic coordinate that we name $x^2$. We say this coordinate is orthogonal to $x^1$ (by construction). 

How do we know that with this we can generate a plane? We consider all possible geodesics among any pair of points belonging to $x^1$ and $x^2$. Any point of these new lines can be characterized by the coordinates $(x^1,x^2)$ that fulfill that the point is at the minimal distance to each axis. It goes without saying that we can repeat this process (generate extra coordinates) as many times as necessary till the whole set of OEs is characterized by the set of geodesic coordinates. At the end of this process we can define {\bf positions} universally: 
\be
{\bf x} \equiv \{x^1,x^2,..., x^d\}. 
\ee
Once furnished with this coordinate representation, one can make standard geometry analyses: angles, ...

{\it Remark}. In this construction, we have implicitly assumed differenciability of the manifold. If we take the cusp of a cone, for instance, the previous discussion is problematic.

\medskip

{\it Transitive property of synchronization}. We can now prove that the synchronization discussed in Sec. \ref{Sincr} holds the transitive property. 

Proof. We consider three points $A$, $B$ and $C$ that lie in a geodesic such that 
$d(A,C)=d(A,B)+d(B,C)$. Since A and B are synchronized we have that $d(A,B)=d(B,A)$. We also have that $d(B,C)=d(C,B)$, since $B$ and $C$ are synchronized. Therefore, we have that 
$d(C,A)=d(C,B)+d(B,A)=d(A,C)$.  This fixes one of the coordinates of the OE. By repeating the process for each coordinate that characterize the OE, we complete the demonstration. 

\subsubsection{Metric space}

Let us first remind the definition of distance and metric space. 

{\bf Definition}. Given any set, $A$, we will say that in $A$ there is defined a {\bf distance} if $\forall x,y, \in A$ we can define a real number,
$d({\bf x},{\bf y})$, with the properties:
\begin{enumerate}
\item
$d({\bf x},{\bf y})=0$ iff ${\bf x}={\bf y}$
\item
$d({\bf x},{\bf y})\leq d({\bf x},{\bf z})+d({\bf y},{\bf z})\qquad \forall {\bf x}$, ${\bf y}$, ${\bf z} \in A$.  
\end{enumerate} 
We name such sets endowed with a distance {\bf metric spaces}.

Using Eq. (\ref{distance}) and synchronization as our definition for distance, any RRF can be considered to be a metric space. Let us prove it.

Our definition satisfies Property 1 by construction. It has to be positive (causality, time always grows) and $d({\bf x},{\bf x})=0$ (definition of OE, if they are different, by definition $d({\bf x},{\bf y})>0$). 

The triangular inequality: 
\be
\label{sim}
d({\bf x},{\bf y}) \leq d({\bf x},{\bf z})+d({\bf z},{\bf y}) \qquad \forall {\bf x}, {\bf y}, {\bf z} \in A\,,
\ee
is also satisfied, otherwise, by this alternative path, we would obtain a shortest distance and then a faster messenger than the limit-messenger, which contradicts the definition of limit-messenger. Therefore, the triangular inequality property our definition of distance has is also consequence of causality. Overall, causality  does not determine the number of dimensions, but  it does determine the mathematical properties that the definition of distance must have. In principle, we do not need $d({\bf x},{\bf y})=d({\bf y}, {\bf x})$. Conceptually they are different things: The first distance refers to the clock of {\bf x} and the second to the clock of {\bf y}. Nevertheless, $d({\bf x},{\bf y})=d({\bf y}, {\bf x})$ happens in our case after synchronization. Overall \underline{all} RRFs, without needing more information/hypotheses, have defined a distance. 

The resulting metric space does not have to be Euclidean (the distance does not necessarily satisfies Pythagoras theorem\footnote{Yes, Pythagoras theorem is not a theorem but an axiom of Euclidean geometry, alternative to the parallel postulate (and indeed more easy to motivate in physical terms).}). If the metric space is Euclidean we have that the distance can be written in the following way: 
\be
d^2({\bf x},{\bf y})=({\bf x}-{\bf y})^2=\sum_{i=1}^d(x^i-y^i)^2
\,,
\ee
when written in terms of the geodesic coordinates.

\subsubsection{Pseudo-metric\footnote{This is not the standard definition of pseudo-metric one finds in mathematical literature. It corresponds to the standard definition if one is restricted to causal intervals. In the physics literature this quantity is also often named metric. }}

It is convenient to define the ($D=d+1$) vector: $x \equiv (x^0,{\bf x})$. The OEs of the RRF are then characterized by world-lines with ${\bf x}$ fixed and changing $x^0$. An event $E$ has associated the point coordinates $x_E=(x^0_E,{\bf x}_E(x^0_E))$. 

We consider the following (Lorentzian) pseudo-metric\footnote{In the following, we omit the subindex $E$ in the coordinate representation of the events.}
\be
d_L^2(x,y) \equiv (x^0-y^0)^2-(d({\bf x},{\bf y}))^2
\,.
\ee
{\it Observation}. $d_L$ is not a metric (but it allows us to define a group). 

\noindent
{\it Observation}. Note that now $x^0-y^0$ does not (necessarily) refers to the minimal time to transfer information between the OE$_{\bf x}$ and the OE$_{\bf y}$. 

Note that, {\it by definition}, we have the following equality (actually identity)
\be
(x^0-y^0)^2-d^2({\bf x},{\bf y})=0
\,,
\ee
if $x^0-y^0$ is the minimal time interval to transfer information between the OE$_{\bf x}$ and the OE$_{\bf y}$, 
and the following inequality
\be
(x^0-y^0)^2-(d({\bf x},{\bf y}))^2>0
\,,
\ee
if $(x^0-y^0)^2=(d^{(m)}({\bf x},{\bf y}))^2$ 
$\forall$ $m$ such that $m$ is not a limit-messenger. Overall, for all possible causally connected events, we always have
\be 
\label{InitialCC}
d_L^2(x,y) \geq 0
\,.
\ee.

It will turn out convenient to cast the above equations in the following form:
\be
\label{finitec}
\frac{d^2({\bf x},{\bf y})}{(x^0-y^0)^2}=1
\,,
\ee
if $m$ is a limit-messenger and 
\be
\label{finitedsm}
\frac{d^2({\bf x},{\bf y})}{(x^0-y^0)^2}<1
\,,
\ee
if $m$ is not a limit-messenger. 

In the first case, we say the vector $x-y$ is a light-like vector, and in the second case that it is a time-like vector. 

{\bf Remark}. Note that Eqs. (\ref{finitec}) and (\ref{finitedsm}) are properties that hold true in any RRF. Indeed, it is also worth emphasizing that the pseudo-metric, $d_L$, we have corresponds to what is named distance in a \underline{synchronous} reference frame (with no time dependence) in general relativity (see \cite{Landau}). 

\subsubsection{Inertial RRF (IRRF)}
{\bf Definition}. We define {\bf IRRF}s as those RRFs for which the Pythagoras theorem holds. In other words, there exists a set of geodesic coordinates of the RRF for which the distance between any two given points ${\bf x}$ and ${\bf y}$ reads
\be
d^2({\bf x},{\bf y})=({\bf x}-{\bf y})^2=\sum_{i=1}^d(x^i-y^i)^2
\,.
\ee
We name these coordinates Cartesian (and this metric Euclidean) because they correspond to the standard definition of Cartesian coordinates if we live in an Euclidean world. 

This metric has associated a pseudo-metric, the "Minkowsky metric" \cite{Minkowski}:
\be
d_M^2({\bf x},{\bf y})\equiv (x^0-y^0)^2-({\bf x}-{\bf y})^2
\,.
\ee

In these RRFs, the conditions (\ref{finitec}) and (\ref{finitedsm}) read
\be
\label{finitecM}
\frac{({\bf x}-{\bf y})^2}{(x^0-y^0)^2}=1
\,,
\ee
if $m$ is a limit-messenger and 
\be
\label{finitedsmM}
\frac{({\bf x}-{\bf y})^2}{(x^{0}-y^{0})^2}<1
\,,
\ee
$\forall$ $m$ such $m$ is not a limit-messenger. 

\subsubsection{The RRFs are smooth manifolds}

We assume continuity/differentiability of the RRF space. Therefore, the RRFs are Riemann manifolds. We can then talk of the concept of "proximity" between the points of the RRF. 

The RRFs are completely characterized by the intrinsic properties of the metric, $d({\bf x},{\bf y})$, following the pioneering work of Gauss in two dimensions, and the general solution for smooth manifolds given by Riemann. For the purposes of this essay, we only make the distinction between Euclidean and Non-Euclidean metrics, and focus on the former. 

We can always chart the space of RRFs with geodesic coordinates (up to global considerations), what happens in noneuclidean spaces is that 
$d^2({\bf x},{\bf y}) \not= (x^1-y^1)^2+(x^2-y^2)^2+(x^3-y^3)^2$ (in $d=3$ dimensions).\footnote{Not of major interest to us but $d=1$ Riemann spaces are always Euclidean.} 
In this context, the real importance of the IRRFs is that they always appear as the short distance limit of smooth manifolds (which is tantamount to say to the short distance limit of all RRFs). At the practical level (in physical processes), these small patches can be considered to be quite big. This is important. Since we know that any Riemann manifold can be organized as small (differential) patches, and for each of them the metric can be approximated to the Euclidean metric, at short distances, we have that the distance (any distance, as far as it is distance) can be written in the following way:
\be
\label{dCart}
d({\bf x},{\bf y})=({\bf x}-{\bf y})^2+{\it o}\left(({\bf x}-{\bf y})^2\right) 
\,.
\ee

If we take Eqs. (\ref{finitec}) and (\ref{finitedsm})
in the short distance limit, we get
\be
\label{infinitesimalc}
\lim_{{\bf x}\rightarrow {\bf y}}\frac{({\bf x}-{\bf y})^2}{(x^0-y^0)^2}=1
\,,
\ee
if $m$ is a limit-messenger, 
and
\be
\label{infinitesimaldsm}
\lim_{{\bf x}\rightarrow {\bf y}}\frac{({\bf x}-{\bf y})^2}{(x^0-y^0)^2}<1
\,,
\ee
$\forall$ $m$ if $m$ is not a limit-messenger. 

These equations are nothing but Eqs. (\ref{finitec}), (\ref{finitedsm}) 
but for a particular (the Euclidean) realization of the metric, i.e. Eqs. (\ref{finitecM}) and (\ref{finitedsmM}). 

\section{Family of IRRFs that are causally connected}

We now consider two events that are causally connected in one RRF$_A$. This means that Eq. (\ref{InitialCC}) holds, where we have characterized these two events by the four vectors $x$ and $y$. This means that there exists a messenger that can transfer the information from the point ${\bf x}$ at the time $x^0$ to the point ${\bf y}$ at the time $y^0$. We now consider a second RRF$_B$, and assume that we can communicate information between the two RRFs, otherwise Occam's principle apply. But then they can be considered messengers between each other. Then, if two events are causally connected in one RRF, they have to be causally connected in the other RRF, otherwise one would enter into a logical contradiction with our construction hypothesis of the RRFs. Let us give two examples:

A) In RRF$_A$ the flashlight emits a signal in ${\bf x}$ at $x^0$ and reaches ${\bf y}$ at $y^0 > x^0$ where it is reflected by a mirror. If in RRF$_B$ the order of events were reversed ($x'^0>y'^0$), we would have that, out of nothing, light would come out of the mirror that would get absorbed by the flashlight.

B) To make option A) more extreme, we can think that in RRF$_A$ an arrow is sent in ${\bf x}$ at $x^0$ and reaches ${\bf y}$ at $y^0 > x^0$ where it kills the observer. If in RRF$_B$ the order of events were reversed ($x'^0>y'^0$), we would have that a dead observer with an arrow stuck in the body would come out into live and the arrow would come out of his body spontaneously in reverse movement till it reaches back the bow. 

Overall, in both cases, the RRF$_B$ would not be a RRF as we have defined it. Therefore, we demand that the allowed set of transformations (automorphisms) between RRFs should fulfill that 
\be
\fbox{
$\displaystyle{
{\rm if}\;\;\; d_L^2(x,y) \geq 0 \quad {\rm then} \quad d_L'^2(x'(x),y'(y)) \geq 0. 
}$
}
\ee

We name this equation the causality condition (between physical events). 

Note that the metric in the RRF' could be different to the metric in RRF: 
\be
d_L'^2(x',y')=(x'^0-y'^0)^2-d'^2({\bf x}',{\bf y}').
\ee
where $d'^2({\bf x}',{\bf y}')$ is the metric in RRF'. 

This far the discussion is general for arbitrary RRFs. We now restrict the discussion to obtain transformations between IRRFs that preserve its Minkowski structure. In other words, we consider two different IRRFs and only allow transformations between them that fulfill the condition
\be
\label{CausalityM}
{\rm if}\;\;\; d_M^2(x,y) \geq 0 \quad {\rm then} \quad d_M^2(x'(x),y'(y)) \geq 0. 
\ee
The solution to this problem was obtained in Refs. \cite{Alexandrov,Zeeman}. In \cite{Zeeman} this was stated as the following theorem:\footnote{We do not explicitly write the proof of this and following theorems in this essay. They can be found in the original papers. Nevertheless, they should be carefully explained when teaching this material. Actually, this also applies to some observations/remarks throughout the text. Some of them could be left as exercises to the students.} 

\noindent
{\bf Theorem 1} \cite{Zeeman} The maximal set of transformations $x'=g(x)$ between IRRFs with $d \geq 2$ that fulfill Condition (\ref{CausalityM}) form a group, which we name the {\it Causal Group $\equiv G$}, where $G$=\{Translations, Rotations, Dilatations, Parity flip, Boosts\}.

{\it Remark}. $G$ is linear: $x'=g(x)=\Lambda x+a$, where $\Lambda$ is a real $D\times D$ matrix. This is a consequence of causality. 

{\it Remark}. Except for parity, $G$ is a continuous group. This a consequence of causality. 
We can then talk of the concept of "proximity" between different IRRFs. We see that the parameters that characterize the transformation between IRRFs can be made very small (leaving aside parity). We can then talk of continuity in this set of parameters. This leads to the appearance of Lie Groups (and the associated Lie algebras) in the characterization of these transformations. The different elements of the group can then be obtained by infinite composition of infinitesimal transformations via exponentiation. The resulting parameters are named normal parameters and fulfill the additive property. 

\medskip

We now discuss in more detail the different transformations that form $G$. We first consider Translations, Rotations, Dilatations and Parity flip.

\begin{itemize}
\item
{\bf Translations}
\be
\label{Translations}
x^{\mu} \longrightarrow x^{\mu}+a^{\mu}
\,,
\ee
where $a^{\mu} \in \mathbb{R}^4$ are the normal parameters for translations. In physical terms, this is nothing but deciding changing the origin of coordinates for time and space. 
\item
{\bf Rotations}
\be
{\bf x} \longrightarrow R{\bf x}
\ee
where $R$ is a real $d\times d$ matrix that fulfils $R^{T}R=I$ and $det(R)=1$. In terms of normal parameters this matrix can be characterized by the angles that define the direction of a vector of modulus 1 in $d$ dimensions. 
\item
{\bf Time (and space) dilatations}
\be
\label{dilatations}
x^{\mu} \longrightarrow e^{\lambda} x^{\mu}
\ee 
where $\lambda \in \mathbb{R}$. 
Note that in our construction, if we change the frequency of the clocks, we also change the rulers with which we measure distance in the same way. Therefore, the causality condition (\ref{CausalityM}) still holds. 
\item
{\bf Reverse parity}
\be
{\bf x} \longrightarrow - {\bf x}
\ee
Strictly speaking we only have to change the sign of an odd number of components of the vector, otherwise it is already included in rotations.
\end{itemize}

The different IRRFs generated by these transformations can be considered to be the same set of OEs with the qualification that the OEs have decided to change their conventions for measuring things. Looked in this way, the would-be different IRRFs would indeed be the same IRRF. This makes self evident the synchronization of clocks between these "different" set of OEs: in the above transformations $x^0=x'^0$ except (obviously) for translations and dilatations, but in these two last cases the sign of time differences do not change either. A complete different thing will be when we consider IRRFs such that the distance among their coordinates changes over time. We discuss them in the next subsection.

\subsection{IRRF generated by messengers (boosts)}
A general boost in terms of the normal coordinates $\boldsymbol{\eta}\equiv\eta \,{\hat n} \in \mathbb{R}^3$, where ${\hat n}=\{n^1,n^2,n^3\}$ is a $d$ dimensional vector of modulus one, reads
\be
\label{boosts}
x^{\mu} \longrightarrow (B({\bf \eta}) x)^{\mu}
\,.\ee 
where
\be
\label{Eq:boostsrapidity}
B({\bf \eta})=
\begin{pmatrix}
\cosh(\eta) & -\sinh(\eta) n^1 & -\sinh(\eta) n^2 & -\sinh(\eta) n^3 \\
-\sinh(\eta) n^1 & 1 + (\cosh(\eta) - 1) (n^1)^2 & (\cosh(\eta) - 1) n^1 n^2 
& (\cosh(\eta) - 1) n^1 n^3 \\
-\sinh(\eta) n^2 & (\cosh(\eta) - 1) n^2 n^1 & 1 + (\cosh(\eta) - 1) (n^2)^2 & (\cosh(\eta) - 1) n^2 n^3 \\
-\sinh(\eta) n^3 & (\cosh(\eta) - 1) n^3 n^1 & (\cosh(\eta) - 1) n^3 n^2 & 1 + (\cosh(\eta) - 1) (n^3)^2
\end{pmatrix}
\,.
\ee
This expression gives the coordinate representation in IRRF': $x'=Bx$ of the event $x$ in the IRRF. If we consider the world-lines associated to the coordinates of the IRRF', we observe that such coordinates move with constant velocity ${\bf v}$ when measured in IRRF. The relation between the normal parameters $\boldsymbol{\eta}$ and ${\bf v}=(v^1,v^2,v^3)$ is the following:
\be
\label{etav}
\sinh(\eta){\hat n}=\frac{1}{\sqrt{1-{\bf v}^2}}{\bf v}
\,.
\ee
In terms of ${\bf v}$, the matrix $B$ reads
\be
\label{Eq:boosts}
B({\bf v})=
\begin{pmatrix}
\gamma & -\gamma v^1 & -\gamma v^2 & -\gamma v^3 \\
-\gamma v^1 & 1 + (\gamma - 1) \frac{(v^1)^2}{|\mathbf{v}|^2} & (\gamma - 1) \frac{v^1 v^2}{|\mathbf{v}|^2} & (\gamma - 1) \frac{v^1 v^3}{|\mathbf{v}|^2} \\
-\gamma v^2 & (\gamma - 1) \frac{v^2 v^1}{|\mathbf{v}|^2} & 1 + (\gamma - 1) \frac{(v^2)^2}{|\mathbf{v}|^2} & (\gamma - 1) \frac{v^2 v^3}{|\mathbf{v}|^2} \\
-\gamma v^3 & (\gamma - 1) \frac{v^3 v^1}{|\mathbf{v}|^2} & (\gamma - 1) \frac{v^3 v^2}{|\mathbf{v}|^2} & 1 + (\gamma - 1) \frac{(v^3)^2}{|\mathbf{v}|^2}
\end{pmatrix}
\,,
\ee
where  
\be
\gamma=\frac{1}{\sqrt{1-{\bf v}^2}}
\,.
\ee

Compared with the transformations in the previous section, boost transformations can be genuinely interpreted as different observers in terms of messengers.  We can use the messengers to generate new IRRFs. By construction, for these, the distance between the messengers and the original OE changes over time (looked from the point of view of the original IRRF). This change is constant over time:
\be
v^{(m)}=\frac{d_{ij}}{d^{(m)}_{ij}} \;,\qquad
\frac{d}{dx_i^0}v^{(m)}=0
\;.
\ee
Therefore, we can do a mapping between these $v^{(m)}$ and the parameters ${\bf v}$ that appear in Eq. (\ref{Eq:boosts}). Obviously, the physical interpretation of these $v^{(m)}$ is that they are the relative velocity of IRRF' with respect to IRRF. Note also that $v^{(m)}$ is nothing but Eq. (\ref{finitedsmM}). This makes evident that $|{\bf v}| \leq 1$. Finally, 
we can also see that the limit-messengers: light/graviton/... are the messengers that give the maximum speed. 

That boost transformations preserve causality is well known. Actually the nontrivial result of Refs. \cite{Alexandrov,Zeeman} is not that $G$ is causal but that there are no more allowed transformations consistent with causality for Minkowski metrics. In this respect, what the present paper yields is a physically motivated path to the initial hypotheses of these works. 

{\it Observation}. Note also that, any two IRRFs, if they are causally connected, move with constant relative velocity. This, which somewhat corresponds to the relativity principle, is a consequence of the present derivation, rather than taken as an hypothesis. 

{\it Observation}. In some derivations of special relativity, the group structure of the coordinate transformations between IRRFs is taken as an hypothesis, whereas here appears as a consequence. 

\subsection{Weaker versions of Theorem 1}
Remarkably enough, in Ref. \cite{Zeeman} weaker versions of Theorem 1 were also presented:\\
{\bf Theorem 2} \cite{Zeeman} The maximal set of transformations between IRRFs with $d \geq 2$ that fulfill the condition: 
\be
\label{CausalityM2}
{\rm if}\;\;\; d_M^2(x,y) > 0 \quad {\rm then} \quad d_M^2(x'(x),y'(y)) > 0. 
\ee
is the {\it Causal Group $\equiv G$}.
 
{\it Corollary of Theorem 2}. $G$ can be obtained in a world with no limit-messenger particles (i.e. in a world with only massive particles). This means that, even if we do not have limit-messengers, all elements of $G$ can be obtained (we are talking of a set of transformations that have group properties). This means that we can always approach to the (limit speed) limit-messenger case as close as we want by composition of elements of the group (i.e. by boosts). 

Another important theorem is the following:\\
{\bf Theorem 3} \cite{Zeeman} Given a transformation $x'=g(x)$ between IRRFs, it fulfills Condition (\ref{CausalityM2}) if and only if it fulfills Condition
\be
\label{CausalityM3}
{\rm if}\;\;\; d_M^2(x,y) = 0 \quad {\rm then} \quad d_M^2(x'(x),y'(y)) = 0. 
\ee

{\it Corollary of Theorem 3}. A limit-messenger in one IRRF is also a limit-messenger in the other IRRF. This implies that a limit-messenger is a IRRF independent concept.

{\it Corollary of Theorem 3}. The causal group $G$ can be determined only demanding causality to be preserved between events related by limit-messengers.
 
\subsection{Extra remarks}

{\it Remark}. Note that we have defined the distance directly proportional to time. Therefore, space and time have the same units. Consequently, our definition of velocity is dimensionless. The fact that we can do that makes explicit that there is nothing fundamental about the specific value of the speed of limit-messengers (light), other than it is different from zero. 

{\it Remark}. Note that we are not demanding $d_M^2(x'(x),y'(y))$ to be invariant under $G$, as it is often done in derivations of special relativity transformation rules, we only demand the sign of $d_M^2(x'(x),y'(y))$ to be invariant for time-like or light-like $x-y$ vectors. Indeed, it is only invariant up to a scale factor. 

{\it Remark}. One obtains the same set of transformations if one only uses causality of limit-messengers (light). In other words, Eq. (\ref{CausalityM3}) is the only condition that has to be preserved by the transformations. One could be worried that this may enter in contradiction with the result that the most general transformation that leads invariant light rays is the conformal group,\footnote{An explicit demonstration can be found, for instance, in Ref. \cite{Fock}.} which is larger than the previous considered group $G$ (and not linear). Nevertheless, the conformal group does not preserve causal ordering. This is due to the fact that special conformal group transformations:
\be
\label{SpecialConformal}
x^{\mu} \longrightarrow \frac{x^{\mu}-b^{\mu}x^2}{1-2b\cdot x +b^2x^2}
\ee
can violate causality. One can easily see this by considering the inversion operation. one can also see this if one considers the finite special conformal transformation formula with $x^0$  unbounded. This implies that the sign of $x'^0$ could be flipped for large $x^0$ even if $b^0$ is small. If one elliminates those special conformal transformations one has again $G$. 

{\it Remark}. The OEs of different IRRFs are causally connected (according to our definition), and will remain them to be so forever. This could be interpreted as a conservation law. 

{\it Remark}. The clocks in IRRF' are synchronized in the standard way, as discussed in Sec. \ref{Sincr}. We could still fix one point (typically the origin) such that $x_0=x_0'=0$, and the axis directions, but no more. Nevertheless, time differences of the clocks in IRRF' have to be compatible with the values obtained after a boost transformation (since IRRF' can be understood to be a messenger generated by a boost). If this does not happen, it means that boost transformations mix with dilatations such that the synchronization of the clocks of the messengers is compatible with the result of the group transformation. In other words, the coordinate change associated to messengers moving with velocity ${\bf v}$ corresponds to an element of the group $G$, where $\boldsymbol{\eta}({\bf v})$ is Eq. (\ref{etav}) but $\lambda({\bf v})$ has a nontrivial dependence in ${\bf v}$. If this happens the space is not isotropic. An example of the lost of isotropy in two dimensions (which in this case is nothing but parity) can be found in \cite{Leblond}. An example in four dimensions can be found in Ref. \cite{Drory}. It is remarkable, still, that even if space is not isotropic, the messenger-limit speed is. 

{\it Remark}. Space-time is homogeneous for IRRFs. 

Finally, we could think of generalizing this discussion to the Non-Euclidean case. We do not consider this possibility here. We will content with the observation that all RRFs are Riemann manifolds. Therefore, we could still apply Alexandrov-Zeeman theorems in its infinitesimal version to all RRFs. 


\section{Conclusions}

There are many ways in which one can obtain that the allowed symmetry group transformations between different inertial reference frames contains the Poincaré group (different set of assumptions yield this or similar results). This gives a feeling of Poincaré symmetry being "unavoidable" for any sensible theory. This triggers seeking the most fundamental\footnote{A concept which is fundamentally ambiguous.}/minimal set of hypotheses. This is obviously of major relevance when trying to present this subject to undergraduate students. Here, as it could not be otherwise, we have followed the causality path.

We have given a construction of (special) relativity based on assuming the impossibility to have instantaneous interactions between observers located at different points in space. In our opinion this is a natural requirement that avoids potential paradoxes one may have otherwise. This requirement leads to the existence of a maximal velocity for transfer of information among different observers. In other words, all messengers between different observers move at velocities smaller or equal than this maximal velocity. Or even restated differently: All messengers need a finite, nonzero time, to reach an observer located in a different position in space. This result holds in any reference frame, inertial or not. Causality also follows from this condition, again in any reference frame. 

A geodesic is usually defined as "the shortest line" (or segment). When one comes to think about it one realizes that we do not really know what a line is. Here we give an operational definition of "shortest line"/geodesic: Minimal transfer information time. 
This definition provides with an experimental construction of geodesic coordinates. Note that this reverses the logic about light: Rather than saying that light travels through the geodesics, we define the geodesics as the path followed by light (limit-messenger). Once we have the geodesic coordinates, we can use them to characterize the position of the OEs in the RRF.

We have then defined IRRFs as those RRFs where there exists geodesic coordinates such that the Pythagoras theorem holds. In this situation, the geodesic coordinates are the Cartesian coordinates. In physical terms, it is interesting to see that this could be taken as a {\it definition} of the RRFs where there are no forces acting on the messengers except in the interaction points (this quantifies the famous statement "the laws of physics take a simpler form" or that "particles move freely between interaction points"). 

Finally, we considered the allowed relations between different IRRFs. The requirement that the transformation between RRFs for which the distance fulfills Pythagoras theorem preserve causality limits these transformations to be the orthochronous unhomogeneous Lorentz (Poincaré) group times dilatations, a result obtained in Refs. \cite{Alexandrov,Zeeman}. Once reached this point, standard results known for special relativity (plus dilatations) follow. 
 
We finish this assay with some few extra remarks. It is worth mentioning that the concept of relativity between RRFs, as such, is not used in the construction made in this paper of the IRRFs nor in the determination of the allowed transformation rules between them. It happens to be a consequence of living in an Euclidean world in space where causality holds. It does not show up either in our construction the speed of light. On the other hand there is always a limit speed, which is the same in all RRFs by construction/definition. Note also that this speed limit is "1" for all RRFs, since space has the same units than time by construction. Indeed, when one thinks of it, one realizes that space is nothing but the measured time intervals of the observer for some specific set of events. 

\bigskip

{\bf Acknowledgments} This work was supported in part by the Spanish Ministry of Science and Innovation (PID2020-112965GB-100 and PID2023-146142NB-100).


\end{document}